\documentstyle[12pt,epsfig]{article}

\newcommand{\be}{\begin{equation}}
\newcommand{\ee}{\end{equation}}
\newcommand{\ba}{\begin{eqnarray}}
\newcommand{\ea}{\end{eqnarray}}
\begin {document}

\begin{center}
{\Large \bf Updating $V_{us}$ from kaon semileptonic decays}\\
\vspace{1.3cm}

{\large G. Calder\'on and G. L\'opez Castro}\\

{\it Departamento de F\'{\i}sica, Cinvestav del IPN \\
Apartado Postal 14-740, 07000 M\'exico D.F. M\'exico}
\end{center}
\vspace{1.3cm}

\begin{abstract}
We update the determination of $|V_{us}|$  using semielectronic and
semimuonic decays of $K$ mesons. A modest improvement of 15\% with
respect to its present value is obtained for the error bar of this matrix
element when we combine the four available semileptonic decays. The
combined effects of long-distance radiative corrections and nonlinear
terms in the vector form factors can decrease the value of $|V_{us}|$ by
up to 1\%.
Refined measurements of the decay widths and slope form factors in the
semimuonic modes and a more accurate calculation of vector form factors
at zero momentum transfer can push the determination of
$|V_{us}|$ at a few of percent level.

\end{abstract}

PACS: 12.15.Hh, 13.20.Eb, 11.30.Hv,13.40.Ks\\

\newpage

\begin{center}
{\bf 1. Introduction}
\end{center}

  $V_{ud}$ and $V_{us}$ are the most accurate entries of the
Cabibbo-Kobayashi-Maskawa (CKM) matrix \cite{ckm} that have been
determined up to now. Their values recommended by the Particle Data Group
are \cite{pdg2000}: 
 \begin{eqnarray}
|V_{ud}| &=& 0.9735 \pm 0.0008\ ,  \\
|V_{us}| &=& 0.2196 \pm 0.0023\ .
\end{eqnarray}
 When they are combined with $|V_{ub}|=0.0036 \pm 0.0010$ \cite{pdg2000},
the most precise test of the unitarity condition of the CKM matrix up to
date becomes:
\begin{equation}
|V_{ud}|^2+|V_{us}|^2+|V_{ub}|^2 =0.9959 \pm 0.0019\ .
\end{equation}
Neither the central value nor the error bar quoted for $|V_{ub}|$ play any
role in the present test of this unitarity condition. The error bars
quoted in Eq. (1) for $V_{ud}$ and $V_{us}$ contribute to 70\% and
30\% of the total uncertainty in Eq. (3), respectively. A direct
inspection
of Eq. (3) would indicate that the present experimental values for these
entries of the CKM matrix, fail to satisfy unitarity by 2.2$\sigma$.
  This problem makes necessary that further efforts are devoted to
investigate the sources of uncertainties that play a role in
the determination of $V_{ud}$ and $V_{us}$ \cite{motiv} at the
level of $10^{-4}$ and $10^{-3}$, respectively. 

The value quoted for $V_{ud}$ in Eq. (1) arises \cite{pdg2000} from the 
average of their values extracted from Superallowed Fermi
Transitions (SFT) in nuclei and from free neutron beta decay. At
present, the error bar in $V_{ud}$ from SFT is still dominated by
different model-dependent calculations of isospin breaking corrections
\cite{hardy}; despite the
fact that isospin breaking corrections in individual SFT's are 
at a few times $10^{-3}$ level, the resulting uncertainty in the weighted
average for $V_{ud}$ is small (5$\times 10^{-4}$!) \cite{hardy} because a
large set of 9 decays are used in their determination.
On the other hand, the determination of $V_{ud}$ from neutron beta decay 
is reaching the $10^{-3}$ accuracy due to recent improvements in the
measurements of the neutron lifetime \cite{neutron} and the ratio of
its vector and axial couplings 	\cite{gagv}. We have
recently reviewed \cite{nos} this determination of $V_{ud}$ by putting
careful attention to the sources of uncertainties in the neutron decay
rate at the $10^{-4}$ level. It was concluded \cite{nos} that present
inconsistencies among the measurements of the axial-vector form factor
$g_A(0)$ \cite{gagv} are behind the main limitations  in order to
have an alternative accurate determination of $V_{ud}$.

  In the present paper, we focus on the determination of $V_{us}$ from
kaon semileptonic decays\footnote{The determination of $V_{us}$ from
semileptonic hyperon decays still suffers of a reliable calculation of
SU(3) symmetry breaking effects \cite{flores}.}. The value quoted in Eq.
(2), was determined in 1984 by Leutwyler and Roos \cite{lr} using kaon
semielectronic decays:
$K \rightarrow \pi e \nu_e$ ($K_{e3}$). Several articles  (see for
example \cite{pest,gus}) and comprehensive review papers have
appeared \cite{bologna,hocker} that make different updates to the value
of $|V_{us}|$ reported in \cite{lr}. Some of them (see for example ref. 
\cite{bologna}), however, combine 
old data for the integrated spectra (those of ref \cite{lr}) with new
information on the decay widths of $K_{e3}$. Since
new information on the decay widths and form factors of $K_{e3}$ decays
has been accumulated since Leutwyler and Roos' original work (which is
based in Ref. \cite{pdg82}), we
would like in this
paper to explore their impact in the determination of $V_{us}$. In
addition, in the present paper we also include in our analysis the data
corresponding to kaon semimuonic decays ($K_{\mu 3}$). We paid particular
attention to the effects of long-distance radiative corrections in the
and of non-linearities in the squared momentum-transfer dependence of the
form factors in the extraction of $|V_{us}|$. It is found that those
combined effects can decrease the central value of $|V_{us}|$ by up to
1\%. We are not able to sensibly improve the accuracy in
the determination of $V_{us}$ with respect to Eq. (2), but we can identify
some elements of the analysis that, if improved, will help to obtain a
more refined and consistent value of this CKM matrix element.

\

\begin{center}
{\bf 2. Decay amplitude and form factors}
\end{center}

  Let us start by defining the tree-level decay amplitude for the $K (p)
\rightarrow \pi(p') l^+(p_1) \nu_l(p_2)$  ($K_{l3}$, with
$l=e,\mu$) decays \cite{bailin}:
\begin{equation}
{\cal M} = \frac{G}{\sqrt{2}} V_{us}C_K \langle
\pi(p')|\bar{u}\gamma_{\mu}
s| K(p) \rangle \bar{v}(p_1)\gamma^{\mu}(1-\gamma_5)u(p_2)\ ,
\end{equation}
where $V_{us}$ is the CKM matrix element we are interested in, $G$ is
the effective weak coupling at the tree-level and
$C_{K^0}=\sqrt{2}C_{K^+}=1$ are flavor isospin  Clebsh-Gordan coefficients
for the hadronic matrix elements.
  The properties of the hadronic matrix elements have been discussed in
numerous papers before (see for example \cite{bailin}). Here we focus on 
some of their properties under flavour symmetry breaking that are relevant
for the determination of $V_{us}$. 

Using Lorentz covariance, we can write the hadronic matrix element
as\footnote{We will also use superindices in the form factors to indicate 
the specific $K\rightarrow \pi$ channel when required.}: \begin{equation}
 \langle \pi(p')|\bar{u}\gamma_{\mu} s| K(p) \rangle = f_+(t)
\left( (p+p')_{\mu}-\frac{m_K^2-m_{\pi}^2}{t} q_{\mu} \right) + f_0(t)
q_{\mu} \left( \frac{m_K^2-m_{\pi}^2}{t} \right)\ .
\end{equation}
The form factors $f_{+,0}$ are Lorentz-invariant functions of the squared
momentum transfer ($t=q^2=(p-p')^2$). They correspond to the $l=1,0$
angular momentum configuration of the $K-\pi$ system in the crossed
channel, respectively.
The kinematical range allowed for the squared momentum transfer in
$K_{l3}$ decays is $ m_l^2  \leq t \leq (m_K-m_{\pi})^2$. The analiticity
of the amplitude for low values of $t$ demands that $f_0(0)=f_+(0)$.

In the limit of exact isospin symmetry, we have ($i=+$ or $0$):
\begin{equation}
f^{K^+ \rightarrow \pi^0}_i(t) = f^{K^0 \rightarrow \pi^-}_i(t).
\end{equation}
This means that the form factors of charged and neutral $K$ mesons should
be equal for all values of $t$ in this limit. If we rewrite the form
factors as follows
\begin{equation}
f_i(t)=f_i(0) \tilde{f}_i(t)
\end{equation}
we have $\tilde{f}_i(0)=1$. 
Thus, isospin symmetry would imply:
\[
f^{K^+ \rightarrow \pi^0}_+(0)=f^{K^0 \rightarrow \pi^-}_+(0)\ ,
 \]  
and, for all values of $t$:
\begin{equation}
\tilde{f}^{K^+\rightarrow \pi^0}_i(t)=\tilde{f}^{K^0\rightarrow
\pi^-}_i(t)\ .
\end{equation}
The effects of isospin and SU(3)-flavour symmetry breaking will modify
these relations. 

The form factors $\tilde{f}^{K\rightarrow \pi} _{+,0}(t)$ have been
measured experimentally \cite{pdg2000} for $K_{l3}$ decays. It has been
found
that a linear parametrization in $t$,
\begin{equation}
\tilde{f}^{K\rightarrow \pi} _{+,0}(t)=1+\frac{\lambda_{+,0}}{m_{\pi}^2}t\
,
\end{equation}
is {\it sufficient} to describe the data, in most of the kinematical range
of $K_{l3}$ decays, to the degree of accuracy attained by experiments.
Note that in 
Eq. (9), the mass scale in the denominator is set by the mass of the
pion emitted in the corresponding $K$ decay. Thus, the isospin symmetry
relation of Eq. (8) indicates that the dimensionful quantities 
$\lambda'_{+,0}\equiv \lambda_{+,0}/m_{\pi}^2$ are the same for charged
and neutral kaon decays.

  For comparison, the experimental results for the slope constants
$\lambda_{+,0}$ as used
in the Leutwyler and Roos' analyses \cite{lr}  and in the present
paper (using the data of Ref. \cite{pdg2000}) are shown in Table 1. We
observe that isospin violation in present data for $\lambda_+$ of $K_{e3}$
and $K_{\mu 3}$ decays are at a few percent
level, as expected. However the average value reported for the $\lambda_0$
slopes in  $K_{\mu 3}$ decays, strongly violates isospin
symmetry \cite{gus}:
 \begin{equation}
b_0=\frac{\lambda'_0(K^0_{\mu 3})-\lambda'_0(K^+_{\mu
3})}{\lambda'_0(K^0_{\mu 3})+\lambda'_0(K^+_{\mu 3})} =0.59\pm 0.37\ .
\end{equation}
This large isospin breaking is usually thought to arise from the small
value of $\lambda_0$ in $K_{\mu 3}^+$ decay\footnote{Note
however that the value $\lambda_0(K^+_{\mu3})=0.0190\pm 0.0064$
measured recently by the KEK-E246 experiment \cite{kek} leads to
$b_0=0.10\pm 0.20$, in better agreement with isospin symmetry. This new
measurement of $\lambda_0$ is compatible with the predictions based on
the Callan-Treiman relation \cite{ct} or with results obtained in chiral
perturbation models \cite{bijnens}.}.
New high precision measurements of $K_{\mu 3}$ decays as those expected
at Da$\phi$ne
will be very useful to understand the  nature of this isospin symmetry
breaking effect.
 In this paper, we will use the values of $\lambda_0$ reported in ref.
\cite{pdg2000} and we will comment on the impact of the new result
reported in ref. \cite{kek} on the determination of $V_{us}$.

  Concerning the value of the form factor at $t=0$, we will use the values
obtained in Ref. \cite{lr} (note however that our numerical results in
sections 4.1--4.2 are obtained using $f_+^{K^0 \rightarrow
\pi^-}(0)=0.9606$, since the value of the $\pi^-$ decay constant used to 
evaluate the chiral corrections is now
0.8\% smaller (see p. 395 in \cite{pdg2000}) than the one used in
\cite{lr}): 
\begin{eqnarray} f_+^{K^0 \rightarrow \pi^-}(0)&=& 0.961\pm
0.008\ , \\
f_+^{K^+ \rightarrow \pi^0}(0)&=& 0.982\pm 0.008\ .
\end{eqnarray}
These values exhibit an small isospin breaking effect:
\begin{equation}
r= \frac{f_+^{K^+ \rightarrow \pi^0}(0)}{f_+^{K^0 \rightarrow \pi^-}(0)}
= 1.022 \ .
\end{equation}
The form factors at $t=0$, Eqs. (11)--(12), incorporate the second order
\cite{ademollogatto} 
SU(3) breaking effects arising from the $s-d,u$ quark mass difference
and the isospin
breaking corrections due to the $\pi^0-\eta$ mixing in the case of
$f_+^{K^+ \rightarrow \pi^0}(0)$ \cite{lr}. By including a full mixing
scheme of neutral mesons $\pi^0-\eta-\eta'$, one would obtain $r\approx
1.026 $ \cite{pest}, to be compared with Eq. (13). In sections 4.1-4.2, we
quote within square brackets our corresponding
results obtained 
using this additional isospin correction.
 The large error assigned to 
Eqs. (11,12) are, at present, the main source of uncertainty in the
value reported in Eq. (2). Other calculations of these
form factors have been done using either the relativistic constituent 
quark model \cite{jaus} or in a formalism based on the Schwinger-Dyson
equations \cite{ds}. Their results are fully consistent with the ones
shown in Eqs. (11,12), but theoretical errors are not provided for them
(Ref. \cite{jaus} provides an error bar  $\delta f_+^{K^0
\rightarrow \pi^-}=^{+0.002}_{-0.006}$ associated to
uncertainties in the strange quark mass). Thus, we will restrict ourselves
to the results quoted in the previous equations.

   The linear parametrizations of the form factors are a convenient
although arbitrary choice to describe experimental data for all values of
$t$  in $K_{l3}$ decays. On the other hand, information about these form
factors can be obtained  from $\tau
\rightarrow K\pi \nu_{\tau}$ decays in the region  $(m_K+m_{\pi})^2 \leq t
\leq m_{\tau}^2$. The vector form factors in $\tau$ decays display a
resonant structure \cite{aleph} such that when they are extrapolated to
low values of $t$ they will naturally induce nonlinear terms. 
Following the results obtained in $\tau \rightarrow \pi\pi \nu_{\tau}$
decays, which suggest that  two or more vector resonances dominate the
2$\pi$ mass distribution \cite{rho}, one can model the vector form factor
in $\tau   
\rightarrow K\pi \nu_{\tau}$ decays as follows \cite{mirkes}:
\begin{equation}
\tilde{f}_+(t)=\frac{1}{1+\beta_{K^*}} [BW_{K^*}(t)+\beta_{K^*}
BW_{K'^*}(t)]\ ,
\end{equation}
where
\[
BW_X(t)=\frac{m_X^2}{m_X^2-t-i\sqrt{t}\Gamma_X(t)\theta(t-(m_K+m_{\pi})^2)}\
.
\]
The subindex $X=K^*(892),K'^*(1410)$ denote the charged vector resonances
in the $l=1$ configuration of the $K\pi$ system, $\Gamma_X(t)$ is the
corresponding decay width and $\beta_{K^*}$ is used
to denote the relative strength of both contributions.
 
When we extrapolate Eq. (14) below the $K\pi$ threshold, we
obtain:
\begin{equation}
\tilde{f}_+(t)=\frac{1}{1+\beta_{K^*}}\left[
\frac{m_{K^*}^2}{m_{K^*}^2-t}+\beta_{K^*} 
\frac{m_{K'^*}^2}{m_{K'^*}^2-t} \right].
\end{equation}
If we set $\beta_{K^*}=0$, and expand the resulting form factor in
powers of $t$, one obtains $\lambda_+=(m_{\pi}/m_{K^*})^2=0.024$. The fact
that a single pole $K^*(892)$ underestimates the value of $\lambda_+$ was
discussed long ago (see for example ref. \cite{yamada}). Now, when we
expand this expression up to terms of order $t$, we can find
the  value of the free parameter $\beta_{K^*}$ from the values of
$\lambda_+$ of each $K_{l3}$ decay according to:
\[
\beta_{K^*}=-\left(\frac{\lambda_+-r_{K^*}}{\lambda_+-r_{K'^*}} \right)\ ,
\]
where $r_X=(m_{\pi}/m_X)^2$. Thus, the vector form factor with two poles
given in Eq. (15) is a natural generalization that includes non-linear
effects in
$t$ and reproduces the correct size of the coefficients in the linear
terms. An estimate of the effects of nonlinear terms in the
determination of $|V_{us}|$ was given for example in ref. \cite{gus}.

\begin{center}
{\bf 3. Decay rates and radiative corrections}
\end{center}

When the radiative corrections are added to the lowest order amplitude,
the decay rate of the $K_{l3}$ decays can be written as follows \cite{lr}:
\begin{equation}
\Gamma(K_{l3}) = \frac{G_F^2m_K^5}{192\pi^3}S_{EW} C_K^2
|f_+^{K\rightarrow \pi}(0).V_{us}|^2 I_K (1+\delta_K)\ ,
\end{equation}
where $G_F=1.16639(1) \times 10^{-5}$ GeV$^{-2}$ \cite{pdg2000} is the
Fermi
coupling constant obtained from $\mu$ decay, and the dimensionless
integrated spectrum $I_K$ is defined as:
\begin{eqnarray}
I_K &=& \frac{1}{m_K^8}\int_{m_l^2}^{(m_K-m_{\pi})^2}\frac{dt}{t^3}
(t-m_l^2)^2\lambda^{1/2}(t,m_K^2,m_{\pi}^2) \nonumber \\ 
&&\times \left\{ \lambda(t,m_K^2,m_{\pi}^2)(2t+m_l^2)|\tilde{f}_+(t)|^2 +
3m_l^2(m_K^2-m_{\pi}^2)^2|\tilde{f}_0(t)|^2 \right\} ,
\end{eqnarray}
where  $\lambda(x,y,z)\equiv x^2+y^2+z^2-2xy-2xz-2yz$. The second term in
$I_K$ (proportional to $m_l^2$) shows that $K_{\mu 3}$  decays are more 
sensitive to the effects of the scalar form factor than $K_{e3}$ decays.
In practice, the factor $(m_K^2-m_{\pi}^2)^2$ in front of the scalar form
factor provides a further suppression for this contribution. This fact
helps our purposes of using $K_{\mu 3}$ decays in our analysis, because
the
issue of the isospin breaking in $\lambda_0$ discussed in the previous
section is not very critical at the present level of accuracy to determine
$V_{us}$.

  As usual, in the above expression for the decay width we have factorized
the radiative corrections into a short-distance electroweak piece 
$S_{EW}$, and a long-distance model-dependent QED correction $\delta_K$.
Since energetic virtual gauge bosons explore the hadronic transitions at
the quark level, $S_{EW}$ is, in good approximation, the same for
all the $K_{l3}$ decays \footnote{In the case of the 
strangeness-conserving
SFT decays the piece of this correction arising from the axial-induced
photonic corrections contributes with one of the important theoretical
uncertainties, $\delta |V_{ud}| \approx 0.0004$ \cite{sirmar86,towner}.
Since the
accuracy 
in the determination of $V_{us}$ does not reach this level yet, we ignore 
here those corrections.}; by including the resummation of
the dominant logarithmic terms one obtains $S_{EW}=1.022$ \cite{sirmar86}.

  The long distance radiative corrections $\delta_K$ are different
for each process since they depend upon the
charges and masses of the particles involved in a given decay. They were
computed long ago for different observables associated to
$K_{l3}$ decays \cite{ginsberg} using the following
approximations: $(i)$ point electromagnetic vertices of the pseudoscalar
mesons, $(ii)$ fixed values of the hadronic weak form factors, and $(iii)$
the local four-fermion weak interaction. The values of $\delta_K$ for each
of the four $K_{l3}$ decays \cite{ginsberg} are shown in the last column
of Table 1. It is interesting to observe that
$\delta_{K}(K^0_{l3})-\delta_{K}(K^+_{l3})\approx 2\%$, both for
semielectronic and semimuonic kaon decays. The origin of this difference
can be traced back to the coulombic interaction between the pion and the
charged lepton in $K^0_{l3}$ decays \cite{ginsberg} (see also
\cite{pest}). We discuss in section 4.2 the
impact of these long-distance radiative corrections in the determination
of $V_{us}$.

\

\begin{center}
{\bf 4. Determination of $V_{us}$}
\end{center}

  In this section we extract the quantities $|f_+(0).V_{us}|$ for each
decay and provide a determination of $|V_{us}|$. We first ignore the
effects of long-distance radiative corrections
in order to compare our results with the ones provided in ref. \cite{lr}.
Later we evaluate the effects of those radiative corrections and comment
on the prospects to improve the accuracy in the determination of $V_{us}$
(see
also refs. \cite{gus,pest}.

\

\begin{center}
{\it 4.1 $|V_{us}|$ without using radiative corrections}
\end{center}
 
  Using the input data of Table 1 and the decay rate of Eq. (16), we
obtain the values for the integrated spectrum $I_K$ and the product
$|f_+(0)V_{us}|$ shown in the second and third columns of Table 2,
respectively. The error bars quoted for the latter quantities include a
1\% uncertainty attributed to long-distance radiative corrections
according to
the prescription\footnote{Since long-distance radiative corrections 
have not been applied to all experimental data used to obtain the average 
values quoted in Table 1 for $K_{l3}$ decay observables, a 1\%
uncertainty is added to the decay widths.} of ref. \cite{lr}. It is
interesting to observe that the values of
$|f_+^{K^+ \rightarrow \pi^0}(0)V_{us}|$ obtained from $K^+_{e3}$ and
$K^+_{\mu 3}$ decays are remarkably consistent among them as demanded by
$e-\mu$ universality, despite the fact that isospin symmetry is strongly
broken in the slope of the scalar form factor\footnote{Should we have
used the result of Ref. \cite{kek} for $\lambda_0$, the quantity
$|f_+^{K^+ \rightarrow \pi^0}V_{us}|$ would have been lowered by almost
1.5\%, destroying the nice agreement with $e-\mu$ universality.}. The
same can be stated (within errors)
for the corresponding quantities in neutral kaon decays. Contrary to ref.
\cite{lr}, we have used the experimental values for the slopes of the form
factors in $K_{e3}$ decays instead of assuming isospin symmetry (namely,
we have not assumed 
$\lambda_+(K^+_{l3})=\lambda_+(K^0_{l3})$ as in \cite{lr}). Observe in 
Table 2 that the value of  $|f_+(0)V_{us}|$ extracted from $K^+_{e3}$ is
almost at the same level of accuracy than in $K^0_{e3}$.

  Now, if we include the isospin breaking corrections to the form factors
at $t=0$ using Eq. (13), we can express the results in Table 2 in terms of  
the quantity $|f_+^{K^0 \rightarrow
\pi^-}(0)V_{us}|$ from the four $K_{l3}$ decays. The different values of
this quantity can be used as a consistency test of the calculations of
the different corrections applied to the semileptonic decays,
namely, this quantity must be the {\it same} for all $K_{l3}$ decays. In
Fig. 1, we plot the values of $|f_+^{K^0 \rightarrow 
\pi^-}(0)V_{us}|$ obtained from the four semileptonic kaon decays. We
observe that their values are consistent among themselves and with their
weighted average
\begin{equation}
|f_+^{K^0 \rightarrow  \pi^-}(0)V_{us}|=0.2113 \pm 0.0010\ [0.2110\pm
0.0010],
\end{equation}
which is displayed as an horizontal band in Fig. 1 (for $r=1.022$). In the
previous 
equation and in the results of this and the following sections, we
show within square brackets the figures corresponding to the choice
$r=1.026$ of the isospin breaking correction (see discussion after eq.
(13)).  The scale factor (see p. 11 of ref. \cite{pdg2000}) associated to
the set of 4 independent measurements of $|f_+^{K^0 \rightarrow
\pi^-}(0)V_{us}|$ is $S=0.41$, indicating a good consistency of those 
results.

  From equations (11) and (18) we obtain the following value\footnote{
If we use only the $K_{e3}$ decays, we would have obtained the weighted
average value $|V_{us}|=0.2196 \pm 0.0022\ [0.2192\pm 0.0022]$, namely 
all the new data on $K_{l3}$ decays accidentally combine to give same
value as in Ref. \cite{lr}.}:
\begin{eqnarray}
|V_{us}|&=&0.2200\times \left(1\pm
\sqrt{(0.0083)^2_{f_+(0)}+(0.0047)^2_{``exp"}}\right)\nonumber  \\
&=& 0.2200 \pm 0.0021 \ [0.2197\pm 0.0021],
\end{eqnarray}
where we have used quotation marks on the experimental error to indicate
that they contain a 1\% uncertainty associated to radiative corrections.
The present uncertainty in $|V_{us}|$ is dominated (75\%) by the
theoretical uncertainty in the calculation of $f_+(0)$ \cite{lr}. Thus,
any experimental effort aiming to improve the accuracy in measurements of
the $K_{l3}$ properties, should be accompanied of an effort to reduce the
error bars in the calculation of form factors at $t=0$.

\

\begin{center}
{\it 4.2 $|V_{us}|$ including  radiative corrections}
\end{center}

  Before we proceed to include the effects of long-distance radiative
corrections in the rates of $K_{l3}$ decays, let us first discuss the
effects of these corrections in the determination of the slope
parameter\footnote{We restrict ourselves to this particular case because
ref. \cite{pdg2000} provides information about the values of the entries
for  $\lambda_+$ obtained with and without radiative corrections effects
in the Dalitz Plot or pion spectrum observables.} $\lambda_+(K^+_{e3})$
from experiments.

  If we use the set of measurements of $\lambda_+(K^+_{e3})$ reported
in \cite{pdg2000} and include the effects of 
radiative corrections in all of them, we obtain the weighted
average $\lambda_+^{rc}=0.0285\pm 0.0019$ (namely, an increase of 2.5\%
with respect to its value in the third column of Table 1).
However, if we evaluate the phase space factor with this corrected value
of $\lambda_+^{rc}$ we obtain $I_K(K^+_{e3})=0.1607$ to be compared with
0.1603 (see Table 1). This is an effect of only 0.2\%, indicating that
attributing a 1\% error bar to the decay rates (see footnote 5) due to
effects of long-distance radiative corrections probably overestimates this
uncertainty.

Thus, we proceed to include explicitly the effects of $\delta_K$ in the
decay rate. Using the input data of Table 1 into Eq. (16), we obtain the
values for the product $|f_+(0)V_{us}|$ shown in the fourth column of
Table
2. Once we include the isospin breaking corrections in the $|f_+(0)|$
values for 
$K^+$ decays, we obtain the following weighted average value from the
four $K_{l3}$ decays:
\begin{equation}
|f_+^{K^0 \rightarrow  \pi^-}(0)V_{us}|=0.2101 \pm 0.0008\ [0.2099
\pm0.0008]\ .
\end{equation}
This quantity is plotted as an horizontal band in Fig. 2 (for $r=1.022$),
together
with the individual values obtained from the four kaon decays after
including isospin breaking corrections from Eq. (13). The
agreement among these four values is equally good (scale factor $S=0.66$)
as in the case where long-distance radiative corrections were excluded
(Fig. 1). Thus, on the basis of the scale factor alone we can conclude
that the set of
4 measurements of the $|f_+^{K^0 \rightarrow  \pi^-}(0)V_{us}|$, obtained  
with and without radiative corrections, provide an equally consistent set
of
data.

Using the average value obtained in Eq. (20) we extract the CKM matrix
element:
\begin{eqnarray}
|V_{us}|_{rc}&=& 0.2187\times \left(1 \pm
\sqrt{(0.0083)^2_{f_+(0)}+(0.0038)^2_{exp}} \right)\nonumber \\
&=& 0.2187\pm 0.0020\ [0.2185\pm 0.0020]\ ,
\end{eqnarray}
which is only 0.6\% smaller than the value in Eq. (19). As in the case of
section 4.1,
the error bar is largely dominated by the
uncertainty in the calculation of $f_+(0)$. For comparison, the
corresponding value obtained from $K_{e3}$ decays alone is
$|V_{us}|=0.2186\pm 0.0021\ [0.2183\pm 0.0020]$. 

  In summary, when we include long-distance radiative corrections in
the decay rates, the value of $|V_{us}|$ decreases by almost 0.6\% and
the error bars remain almost the same. If we compare our result for
$|V_{us}|$ in
Eq. (21) with the one obtained in reference \cite{lr}, Eq. (2), we observe
that the overall uncertainty is being reduced by 15\%. From Eqs. (19) and
(21) we
conclude that any experimental effort aiming to improve the precision in
measurements of $K_{l3}$ properties would not have a significant impact 
on the determination of $|V_{us}|$. A reassessment of the SU(3) breaking
effects in $f_+(0)$ is compelling to attain a greater accuracy in
$|V_{us}|$. 

  However, an improvement in measurements of the properties of
$K_{l3}$ decays would help to assess the requirement of
long-distance radiative corrections. In particular, a consistency check of
these calculations can be provided by verifying that 
the quantities $|f_+^{K^0 \rightarrow  \pi^-}(0)V_{us}|$ are the same in
all four $K_{l3}$ decays. This quantity plays a similar role as the ${\cal
F}t$ parameter in SFT, which must be the same for all the $0^+ \rightarrow
0^+$ nuclear transitions after removing (process-dependent) isospin
breaking and radiative corrections from the $ft$ value of each decay (see
\cite{hardy}).

\

\begin{center}
{\it 4.3 Effects of non-linear form factors}
\end{center}

  In this section we study the effects on the determination of $|V_{us}|$
due to nonlinear terms that could
be present in the vector form factors $f_+(t)$. These nonlinear terms are
naturally induced when we extrapolate the vector form factor measured in
the resonance region to energies below the threshold for $K\pi$
production (see discusion in section 2).

  The strength $\beta_{K^*}$ of the relative contributions of the two
resonances in the model of Eq. (15) can be fixed either, ($i$) from the
slope of the  $\tilde{f}_+(t)$ form factor at low momentum transfer or,
($ii$) from the decay rate
of $\tau \rightarrow K\pi\nu$ decays. Using the first method, we can find
the values of $\beta_{K^*}$ using the expression given after Eq. (15).
The values obtained in this way are shown in the second column of Table 3.
These values of
$\beta_{K^*}$ are small and negative as expected
from SU(3) symmetry\footnote{In a vector dominance model we would expect 
$\beta_{K^*} \sim g_{K'^*K\pi}f_{K^*}/(g_{K^*K\pi}f_{K'^*})$ and a
similar expression for $\beta_{\rho}$ with $K^*(K'^*)$ and $K$ replaced
by $\rho(\rho')$ and $\pi$, respectively. Using SU(3) symmetry one can
relate both $\beta_V$ constants and expect an equality of their magnitudes 
within roughly a 40\%.} considerations, since the corresponding
parameter $\beta_{\rho}$ measured in $\tau \rightarrow \pi\pi \nu$ decays
is also small and negative \cite{rho}.

  The corresponding integrated spectrum factor $I_K$ computed by using
Eqs. (15) and (17) are shown in the third column of Table 3. A comparison
of these
results and  the values of $I_K$ found for the linear case (second column
in Table 2) indicates that in the nonlinear case, the values are shifted
upwards by around 1\% .
Consequently, these nonlinearities in $t$ would decrease the individual
values of  $|f_+(0)V_{us}|$ (and of $|V_{us}|$) by an amount of 0.5\%
(see Table 4).  
Thus, instead of quoting a
value of $|V_{us}|$ in this case, we would like to stress that nonlinear
effects in the vector form factors of $K_{l3}$ decays would be very
important in the precise determination of this CKM matrix element. In this
case, more refined measurements of the this form factor both from the
Dalitz plot or $\pi$ spectrum of $K_{l3}$ decays would be suitable.

\

\begin{center}
{\bf 5. Conclusions}
\end{center}

  In this paper we have used the updated information on semileptonic
decay properties of kaons to determine the $|V_{us}|$ entry of the CKM
mixing matrix. We have employed both, the semielectronic ($K_{e3}$) and
semimuonic ($K_{\mu 3}$) decays of charged and neutral kaons. In addition
to the original work of Ref. \cite{lr}, we have explicitly included the
effects of long-distance radiative corrections in our analysis and have
studied the impact of non-linear vector form factors. 

  Our results are summarized in Table 4. We observe that the determination
of $|V_{us}|$ from the semielectronic, the semimuonic and from the
combined modes are
consistent among them. The values of $|V_{us}|$ obtained from the muonic
modes are larger than the ones obtained using the semielectronic modes,
although they are less accurate. This difference becomes
smaller when long-distance radiative corrections are included 
in the decay widths. On the other hand, long-distance radiative
corrections tend to reduce the values of $|V_{us}|$ by a 0.3\% (0.7\%) in
the electronic (muonic) channel. The error bars in the case of
the semimuonic channels are still dominated by the experimental
uncertainties in the decay widths and form factor slopes, while the
corresponding error bars from the semielectronic modes are
largely dominated by the theoretical uncertainty in the calculation of the
form factors at zero momentum transfer. When we combine all the four
$K_{l3}$ decay channels, we obtain a determination of $|V_{us}|$ which 
modestly improve the accuracy obtained by ref. \cite{lr}, Eq. (2).

  Concerning the effects of nonlinear form factors at low momentum
transfer, we have considered a vector dominance model with two resonances,
which turns out to be adequate in the resonance region. We fix the
relative contributions of the two resonances by matching the form factor
with experimental values at low energies. The overall effect of
the nonlinear terms is to reduce the value of $|V_{us}|$ by a 0.5\%. Thus,
the combined effect of long-distance radiative corrections and nonlinear
form factors could decrease the value of $|V_{us}|$ by up to 1\%. New
measurements of the vector form factors of $K_{l3}$ decays, particularly
their energy dependence for soft pions (large values of the momentum
transfer), will be very useful to improve the determination of
$|V_{us}|$.

  Finally, we would like to stress that the set of four kaon semileptonic 
decays turns out to be very useful to make a consistency test of the
measurements and the different corrections applied to the decay rates.
In particular, we mean that when isospin breaking corrections are removed
from the vector form factors at zero momentum transfer, we can extract the
product $|f_+^{K^0\rightarrow \pi^-}(0)V_{us}|$ which must be the {\it
same} for all the four $K_{l3}$ decays. In other words, this parameter
plays the same role as the process-independent ${\cal F}t$ values used in
Superallowed Fermi nuclear Transitions to determine $|V_{ud}|$.

\

{\bf Acknowledgements} The authors acknowledge the partial financial
support from Conacyt (M\'exico) under contracts 32429-E and 35792-E.

\newpage

\

\begin{center}
\begin{tabular}{|c|c|c||c|}
\hline\hline 
Experiment & PDG 1982 & PDG 2001 & Ref. \cite{ginsberg}\\
\hline
$K^+_{e3}$ & & & \\
$\Gamma$  & 2.5645$\pm$0.0271  & 2.5616 $\pm$
0.0323 & 
\\
$\lambda_+$ & 0.029 $\pm$ 0.004 & 0.0278 $\pm$ 0.0019 & \\
$\delta_K$(\%) & - & - & --0.45 \\ 
\hline
$K^0_{e3}$ & & & \\
$\Gamma$  & 4.9147$\pm$0.0740  & 4.9385 $\pm$
0.0446 & 
\\
$\lambda_+$ & 0.0300 $\pm$ 0.0016 & 0.0290 $\pm$ 0.0016 & \\ 
$\delta_K$(\%) & - & - & 1.5 \\ 
\hline
$K^+_{\mu 3}$ & & & \\
$\Gamma$  &1.7026$\pm$0.0480  & 1.6847 $\pm$ 0.0426 &  \\
$\lambda_+$ & 0.026 $\pm$ 0.008 & 0.031 $\pm$ 0.008 & \\
$\lambda_0$ & --0.003 $\pm$ 0.007 & 0.006 $\pm$ 0.007 & \\
$\delta_K$(\%) & - & - & --0.06 \\ 
\hline
$K^0_{\mu 3}$ & & & \\
$\Gamma$ &3.4415 $\pm$ 0.0573 & 3.4604 $\pm$ 0.0416 & \\
$\lambda_+$ & 0.034 $\pm$ 0.006 & 0.034 $\pm$ 0.005 & \\
$\lambda_0$ & 0.020 $\pm$ 0.007 & 0.025 $\pm$ 0.006 & \\ 
$\delta_K$(\%) & - & - & 2.02 \\ 
\hline
\end{tabular}
\end{center}
Table 1: Comparison of observables for $K_{l3}$ decays as reported by
the Particle Data Group in 1982 (ref. \cite{pdg82}) and 2001 (ref.
\cite{pdg2000}). The decay
widths $\Gamma$ are given in units of $10^{-15}$ MeV. The last column
displays the long-distance radiative corrections to the decay
widths according to Ref. \cite{ginsberg}.

\newpage

\

\

\begin{center}
\begin{tabular}{|c|c|c|c|}
\hline\hline 
Process & $I_K$ & $|f_+(0)V_{us}|$ (worc) &$|f_+(0)V_{us}|$ (wrc)\\
\hline
$K^+_{e3}$ & 0.1603$\pm$0.0011& 0.2158$\pm$0.0019& 0.2160$\pm$0.0016 \\
$K^0_{e3}$ & 0.1555$\pm$ 0.0008&0.2108$\pm$0.0015& 0.2093$\pm$0.0011 \\
$K^+_{\mu 3}$ &0.1054$\pm$0.0033 &0.2159$\pm$0.0041 & 0.2159$\pm$0.0043 \\
$K^0_{\mu 3}$ &0.1068$\pm$0.0021 &0.2130$\pm$0.0027 & 0.2109$\pm$0.0024
\\
\hline
\end{tabular}
\end{center}

Table 2: Integrated spectrum $I_K$ of $K_{l3}$ decays and values extracted
for the product $|f_+(0)V_{us}|$ with (wrc) and without (worc) radiative
corrections, using the input data from ref. \cite{pdg2000}. 

\

\

\begin{center}
\begin{tabular}{|c|c|c|c|}
\hline\hline 
Process &$\beta_{K^*}$ &$I_K$ & $|f_+(0)V_{us}|$ (worc) \\
\hline
$K^+_{e3}$ & -0.1827 &0.1616$\pm$0.0012& 0.2149$\pm$0.0019 \\
$K^0_{e3}$ &-0.3057  &0.1568$\pm$ 0.0009&0.2100$\pm$0.0016 \\
$K^+_{\mu 3}$ &-0.3060& 0.1065$\pm$0.0049 &0.2147$\pm$0.0054 \\
$K^0_{\mu 3}$ &-0.4454 &0.1079$\pm$0.0032 &0.2119$\pm$0.0036 \\
\hline
\end{tabular}
\end{center}

Table 3: Integrated spectrum $I_K$ of $K_{l3}$ decays and values extracted
for the product $|f_+(0)V_{us}|$ and without (worc) radiative corrections
 using the nonlinear form factors of Eq. (15).

\newpage

\begin{center}
\begin{tabular}{|c|c|c|c|c|}
\hline\hline 
$r$ & Source &$|V_{us}|$ (without r.c.) & $|V_{us}|$ (with r.c.) &
$|V_{us}|$ (without r.c.)\\
 & & linear f.f. & linear f.f. &nonlinear f.f. \\
\hline
 & $K_{e3}$ & 0.2196 $\pm$ 0.0022 & 0.2186 $\pm$ 0.0021 & 0.2187 $\pm$
0.0022 \\
1.022 & $K_{\mu 3}$ & 0.2212 $\pm$ 0.0030 & 0.2196 $\pm$ 0.0029 & 0.2200 
$\pm$ 0.0036 \\
 & All decays & 0.2200 $\pm$ 0.0021 & 0.2187 $\pm$ 0.0020 & 0.2189 $\pm$ 
0.0021 \\
\hline
 & $K_{e3}$ & 0.2192$ \pm$ 0.0022 & 0.2183 $\pm$ 0.0020 & 0.2184 $\pm$
0.0022 \\
1.026 & $K_{\mu 3}$ & 0.2209 $\pm$ 0.0030 & 0.2194 $\pm$ 0.0029 & 0.2198 
$\pm$ 0.0036 \\
 & All decays & 0.2197 $\pm$ 0.0021 & 0.2185 $\pm$ 0.0020 & 0.2186 $\pm
$ 0.0021 \\
\hline
\end{tabular}
\end{center}

Table 4: Values of $|V_{us}|$ extracted for two values of the
isospin-breaking parameter $r$ (see section 2), from different
combinations of
$K_{l3}$ decays and including (fourth column) or not (third column) the
long-distance radiative corrections. Also shown are the values obtained
using non-linear form factors but excluding rad. corrections (fifth
column).

\newpage

\

\vspace{3in}

\begin{center}
\begin{figure}[h]
\label{fig1}
\centerline{\epsfig{file=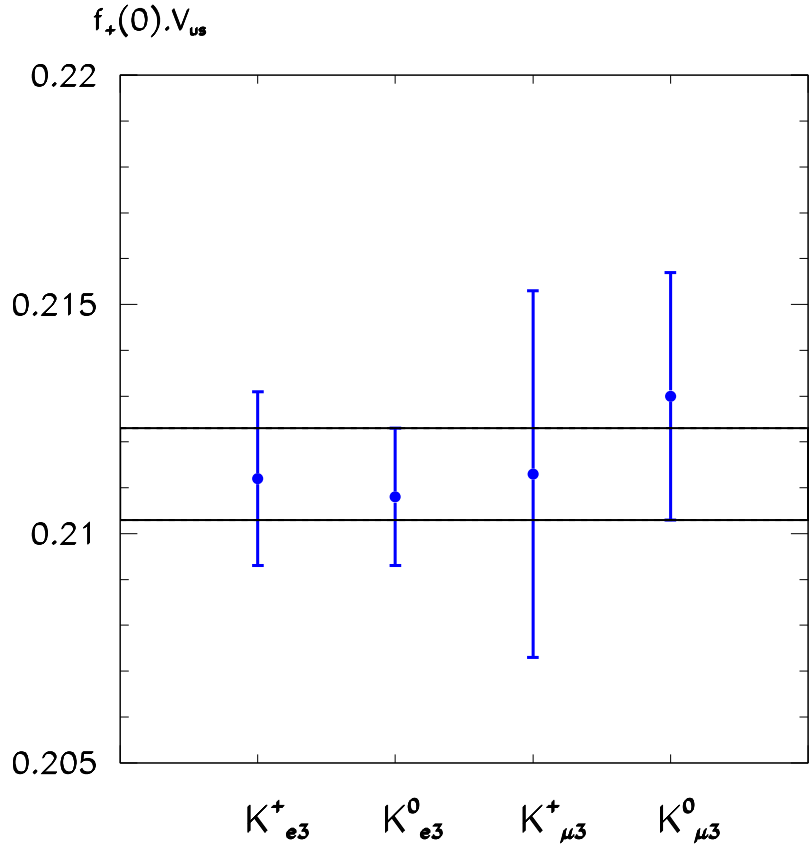,angle=0,width=6in}}
\vspace{-0.1in}
\caption{Values of $|f_+^{K^0 \rightarrow \pi^-}(0).V_{us}|$ obtained
from all $K_{l3}$ decays by ignoring long-distance radiative corrections.
The horizontal band is the 1$\sigma$ weighted average of the four values.}
 \end{figure}
\end{center}

\newpage

\

\vspace{3in}

\begin{center}
\begin{figure}[h]
\label{fig2}
\centerline{\epsfig{file=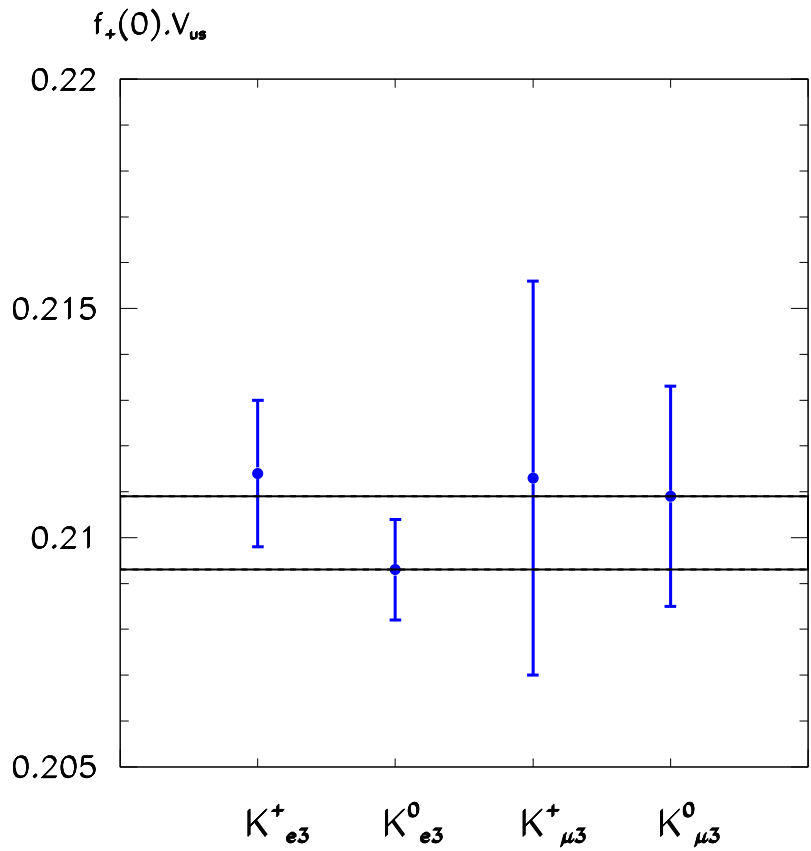,angle=0,width=6in}}
\vspace{-0.1in}
\caption{Same description as in Fig. 1 but when long-distance
radiative corrections are included in the decay widths.}
\end{figure}
\end{center}

\end{document}